\newcommand{\SSS}{\scriptscriptstyle}
\newcommand{\Rl}{R_{\SSS L1}}

\newcommand{\Msol}{{\cal M}_\odot}
\newcommand{\Rsol}{R_\odot}
\newcommand{\Mwd}{M_{\SSS wd}}
\newcommand{\Rwd}{R_{\SSS wd}}
\newcommand{\Pspin}{P_{\SSS spin}}
\newcommand{\Porb}{P_{\SSS orb}}

\newcommand{\lsim}{{\textstyle{\; \lower 0.7ex\hbox{$<$}\;
  \atop \raise-0.1ex\hbox{$\sim$}}}}

\documentclass[onecolumn]{aa}
\usepackage{graphicx}
\usepackage{txfonts}
\begin{document}
\twocolumn
   \title{Is T Leonis a superoutbursting intermediate polar?}
   \authorrunning{Vrielmann, Ness \& Schmitt}
   \titlerunning{}

   \author{Sonja Vrielmann,
	  Jan-Uwe Ness,
          \and
          J\"urgen H.\ M.\ M.\ Schmitt
          }

   \offprints{S. Vrielmann}

   \institute{Hamburger Sternwarte, Universit\"at Hamburg,
	Gojenbergsweg 112, 21029 Hamburg\\
	\email{svrielmann@hs.uni-hamburg.de, jness@hs.uni-hamburg.de,
	jschmitt@hs.uni-hamburg.de}
             }

   \date{Received; accepted}

   \abstract{We present an {\it XMM-Newton} analysis of the
   cataclysmic variable T~Leo. The X-ray light curve shows sinusoidal
   variation on a period $P_x$ equal to $0.89^{+0.14}_{-0.10}$ times
   the previously spectroscopically determined orbital
   period. Furthermore, we find a signal in the power spectrum at
   414~sec that could be attributed to the spin period of the white
   dwarf. If true, T~Leo would be the first confirmed superoutbursting
   intermediate polar (IP).  The {\it spin profile} is double-peaked
   with a peak separation of about 1/3 spin phases. This appears to be
   a typical feature for IPs with a small magnetic field and fast
   white dwarf rotation.

   An alternative explanation is that the 414~sec signal is a
   Quasi-periodic Oscillation (QPO) that is caused by mass transfer
   variation from the secondary, a bright region (``blob'') rotating
   in the disc at a radius of approximately~9$\Rwd$ or -- more
   likely -- a travelling wave close to the inner disc edge of a dwarf
   nova with a low field white dwarf.

   The {\it XMM-Newton} RGS spectra reveal {\it double peaked
   emission} for the O~VIII Ly $\alpha$ line. Scenarios in the IP and
   dwarf nova model are discussed (an emitting ring in the disc,
   bright X-ray spot on disc edge, or emitting accretion funnels), but
   the intermediate polar model is favoured. Supported is this idea by
   the finding that only the red peak appears to be shifted and the
   `blue' peak is compatible with the rest wavelength.  The red peak
   thus is caused by emission from the northern accretion spot when it
   faces the observer. Instead, the peak at the rest wavelength is
   caused when the southern accretion funnel is visible just on the
   lower edge of the white dwarf -- with the velocity of the accreting
   material being perpendicular to the line of sight.
 
   \keywords{stars: binaries: close, stars: novae, cataclysmic
               variables, stars: individual: T Leo, X-rays: stars,
               accretion, accretion discs
               }
   }

   \maketitle
%
\section{Introduction}
Cataclysmic variables (CVs) are close, interacting binaries,
consisting of a white dwarf receiving matter from its companion,
usually -- and so in our case -- a main sequence star. In
non-magnetic systems, the matter is transfered via an accretion disc,
while, if the white dwarf has a significant magnetic field, the inner part
of the disc is disrupted (intermediate polar) or the formation of a
disc is prevented altogether (polar). In both cases the transfered
material couples to the magnetic field lines before falling onto
the white dwarf. CVs have been found to emit radiation across the
whole electromagnetic spectrum, as various energy sources are present
within the system.

This paper focuses on {\it XMM-Newton} X-ray observations of the dwarf
nova T~Leo, i.e., we are looking at high energy sources within the
system. These are usually the boundary layer (e.g., Wood et al.\
1995), a hot corona of the white dwarf (Mahasena \& Osaki 1999), or
accretion funnels and accretion spots on the surface of the white
dwarf in magnetic systems (e.g., Schwarz et al.\ 2002).

   \begin{figure*}
   \centering
   \includegraphics[width=8cm]{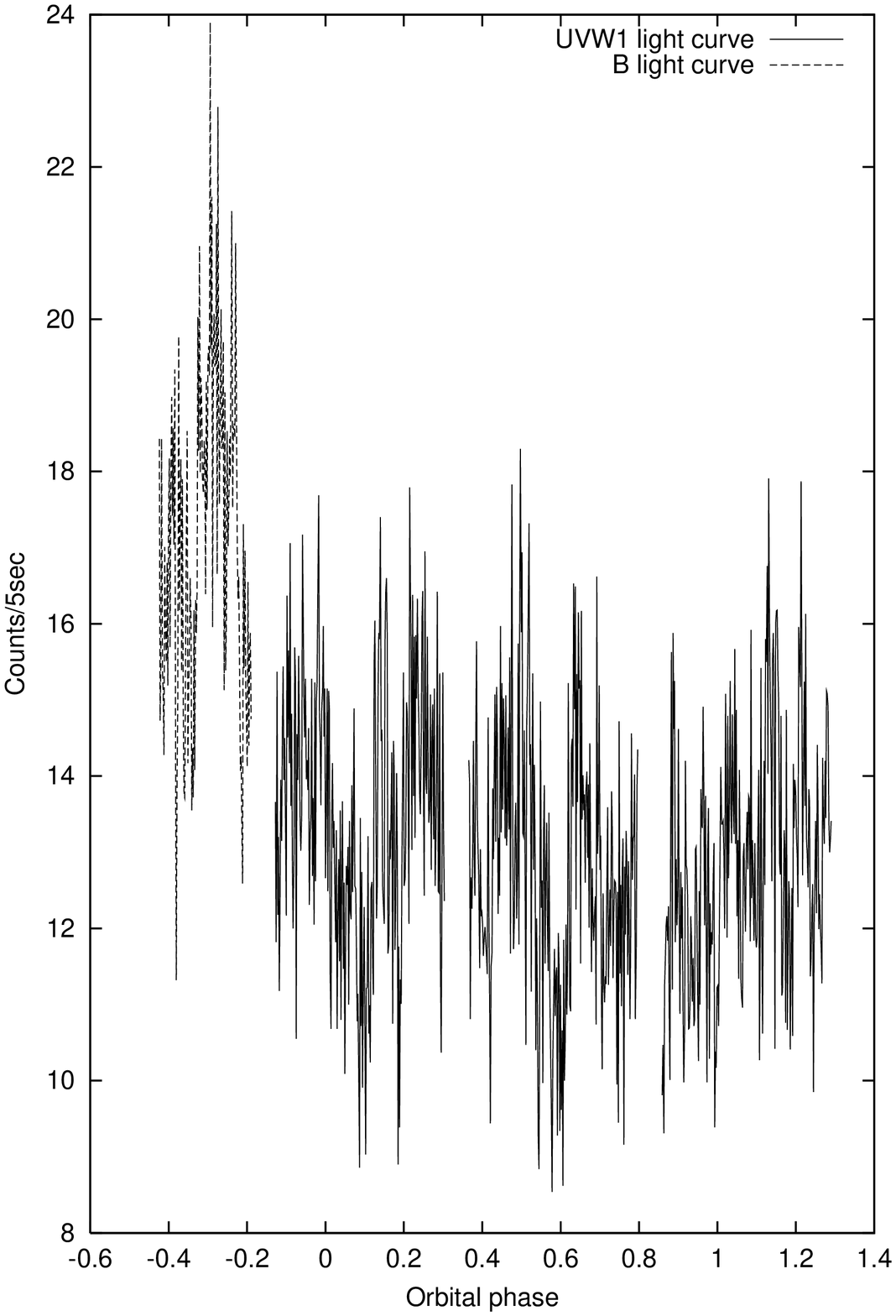}
   \includegraphics[width=8cm]{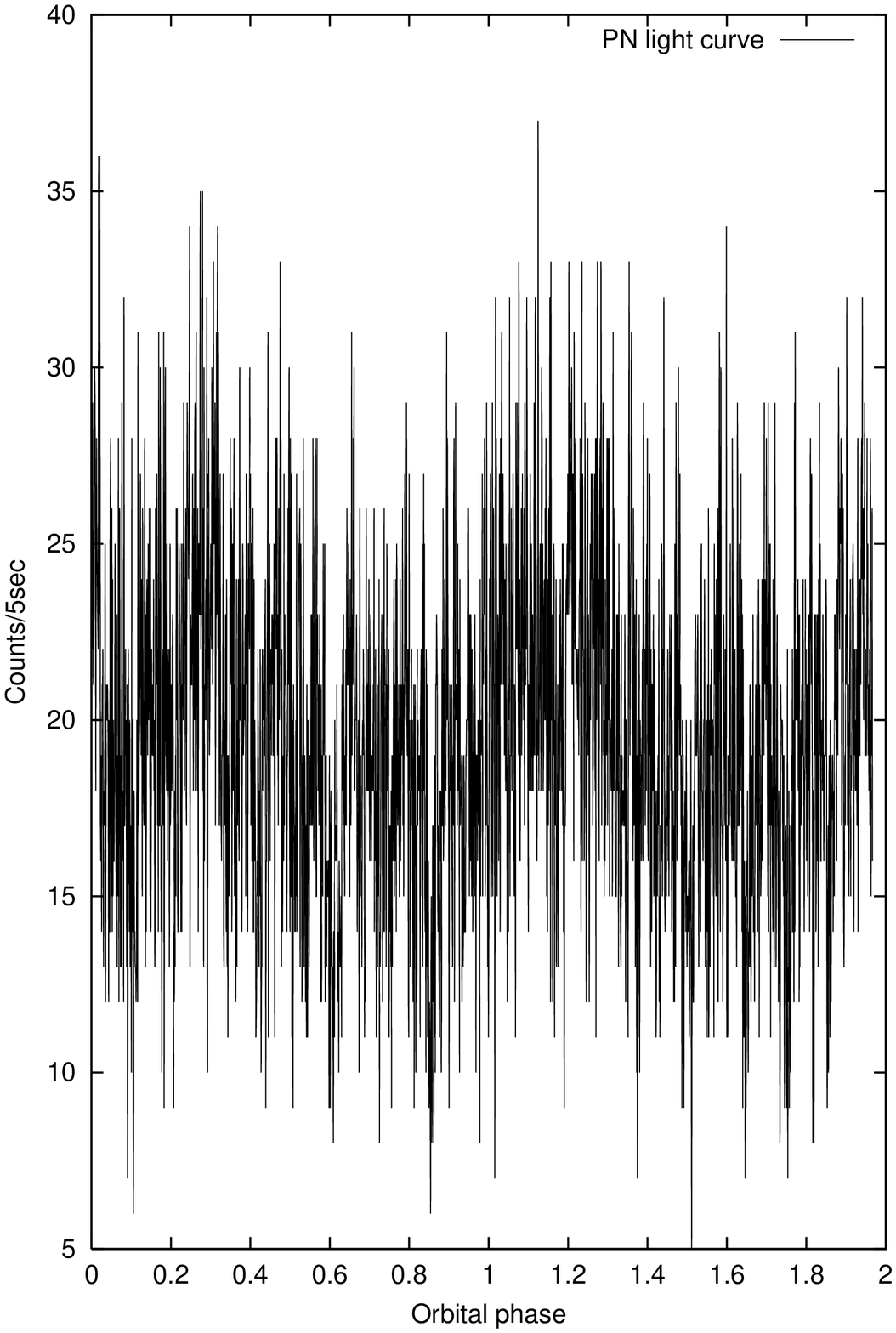}

      \caption{OM (B and UVW1, left) and PN light curves at original
              resolution phased on the orbital period of
	      84~min. (Note, the data set is nearly exactly 2 orbital
	      periods long, no repetition of data is plotted).
              }
         \label{Figpnbom}
   \end{figure*}
%
%

T~Leo is a very short period dwarf nova of SU~UMa type showing
superoutbursts (Slovak, Nelson \& Shafter 1987, Kato \& Fujino
1987). With $\Porb = 84$ min (Shafter \& Szkody 1984) the orbital
period is in fact close to the period minimum of about 76 min (e.g.,
Thorstensen et al.~2002) and makes T~Leo therefore particularly
interesting. So far, T~Leo was only studied twice in detail in X-rays,
once in superoutburst (with {\it RXTE} by Howell et al.\ 1999) and
once in quiescence (with {\it ASCA} by Szkody et al.\ 2001).  Howell
et al.\ detected T~Leo only on one of five days during the outburst
and on that day the X-ray emission was five times lower than during
quiescence (compared with previous pointed {\it ROSAT} PSPC
observations by van Teeseling et al.\ 1996 and Richman 1996). The
timing of the detection coincides with a re-brightening of the system.

It seems to be a typical behaviour of dwarf novae that hard (3-12.2
keV) X-rays are suppressed during outburst as McGowan, Priedhorsky \&
Trudolyubov (2003) find also for SS~Cyg. They suggest that this lack of
hard X-rays, originating in an optically thin disc corona, is caused
by radiation driven disc winds stripping the coronal layers from the
disc during an outburst (Meyer \& Meyer-Hofmeister 1994).

Szkody et al.\ (2001) find a variation on the orbital period for their
quiescent {\it ASCA} X-ray data: just before upper
conjunction of the white dwarf the system brightens
slightly. Otherwise the X-ray emission appears constant.

In this study we investigate spectral properties and the variability
in the {\it XMM-Newton} data of the quiescent T~Leo in order to learn
more about the source of the X-ray emission. Hereby, two basic models
are discussed to explain the behaviour of T~Leo: the intermediate
polar model, as previously suggested by Shafter \& Szkody (1984) or
the dwarf nova model as otherwise assumed (e.g., Kato 1997).


\section{Data}

   \begin{table}
      \caption[]{Observations log. All observations were taken on 1
      June 2002.
      }
         \label{Tabobslog}

         \begin{tabular}{lllll}
            \noalign{\smallskip}
            \hline
            \noalign{\smallskip}
            Instrument & Data Mode &
            Filter & Start UT & Stop UT\\
            \noalign{\smallskip}
            \hline
            \noalign{\smallskip}
 M1  &    Imaging          &   Medium &  11:12:29 & 14:37:11 \\

 M2  &    Imaging          &   Thin1  &  11:12:43 & 14:37:12 \\\hline

 OM  &    Imaging \& Fast   &   B      &  11:09:57 & 11:29:57 \\
 OM  &    Imaging \& Fast   &   UVW1   &  11:35:02 & 12:11:43 \\
 OM  &    Imaging \& Fast   &   UVW1   &  12:16:49 & 12:53:29 \\
 OM  &    Imaging \& Fast   &   UVW1   &  12:58:37 & 13:35:16 \\\hline

 PN  &    Imaging          &   Medium &  11:45:48 & 14:32:28 \\\hline

 R1  &    Spectroscopy     &          &  11:06:11 & 14:40:37 \\

 R2  &    Spectroscopy     &          &  11:06:11 & 14:40:36 \\

            \noalign{\smallskip}
            \hline
         \end{tabular}

   \end{table}

The {\it XMM-Newton} data were taken on 1 June 2002 for a total of
11.4 ksec during which all instruments were used, i.e., OM, EPIC (PN,
MOS1, MOS2), and RGS.  The OM data were taken in the B filter
(duration 1.2 ksec) and UVW1 filter (duration 7.2 ksec). The effective
wavelengths of the filters are 434 nm and 291 nm, respectively. The B
filter data were taken prior to the RGS data, the UVW1 data mostly
simultaneously with the X-ray data (see Table~\ref{Tabobslog}).

For the reduction we used the standard {\em SAS} software (version
5.4.1) without major problems, except for the OM data where we used
the pipeline products.  T Leo is weak enough in X-rays, so no measures
had to be taken to work around pile-up.

\section{Photometry}

Fig.~\ref{Figpnbom} shows the light curves in the optical (B and UVW1
filter) and X-ray range phased on the orbital period $\Porb =
1.41166$~h determined through radial velocity curves in the wings of
the H$\alpha$ emission line (Shafter \& Szkody 1984).  The phasing of
the light curve in Fig.~\ref{Figpnbom} is arbitrary in that phase 0 is
set to the start of the PN light curve.

The light curves display strong flickering hidden in a high noise level as
well as a strong variation dissimilar to that found by Szkody et al.\
(2001). While their light curves brighten only for 20\% of the orbit,
our light curves show rather sinusoidal or saw tooth like variation.

   \begin{figure}
   \centering
   \includegraphics[width=8cm]{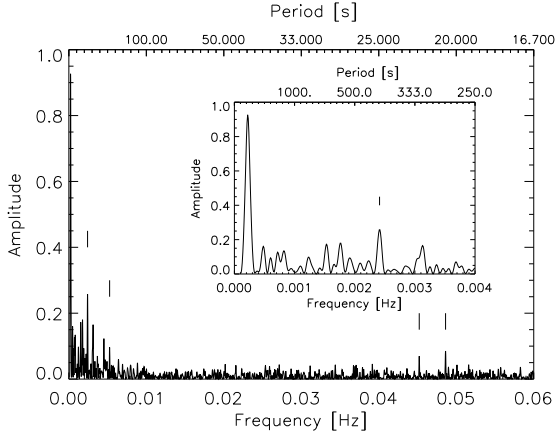}
      \caption{Power spectrum of the PN light curve. The inset is a
      blow up of the low frequency range. The peak at low
      frequencies (at frequency $2\times10^{-4}$ Hz) is consistent
      with the orbital
      period, however, allows a range of 1-1.6 hours. The frequency
      range covers 3~h (length of data set) to 16~s.
      The vertical dashes denote the frequencies cited in the text,
      with periods (left to right) 414.42~s, 190~s, 22.112~s,
      20.567~s. In the inset only the 414.42~s peak is marked.
              }
         \label{Figpower}
   \end{figure}
%

   \begin{figure}
   \centering
   \includegraphics[width=8cm]{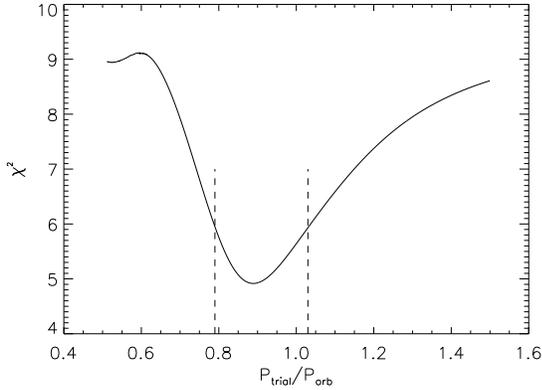}
      \caption{The trial period in relation to the orbital period vs.\
              the $\chi^2$ of the sinusoidal fit. The minimum of the
              $\chi^2$ lies at a period of $0.89 \Porb$. The dashed
	      lines indicate the 1-$\sigma$ error range.
              }
         \label{Figchi2}
   \end{figure}
%

An FFT analysis does, unfortunately, not help in identifying, whether
the variation is linked to $\Porb$, as the length of the data set of
approximately 2 $\Porb$ gives us a very poor resolution in the power
spectrum around $\Porb$ (Fig.~\ref{Figpower}). The width of the peak
around $\Porb$ allows values between 1~h and 1.6~h.

Apart from the orbital peak we can identify a peak in the power
spectrum at $(2.413 \pm 0.04)\times10^{-3}$~Hz (= $414\pm 7$~s) and
its overtones ($2f,3f,4f,5f$). This
could be the spin period of the white dwarf or a Quasi-periodic
oscillation (QPO). Furthermore, we would like to draw attention to the
two minor peaks at 0.049 and 0.045 Hz (20.6~s and 22.1~s). The
discussion of these peaks will follow in Section~\ref{QPO}.

In order to check how stable the peaks are, we split up the PN light
curve into two sets of equal length and analysed both independent sets
individually. The 414~s peak appears in both power spectra. The power
spectrum of the first half of the PN light curve shows a strong series of
peaks of base $f = 1.53\times10{-3}$ Hz (654~s) and its overtones
($2f,3f,5f,6f$, but not $4f$). The first three peaks (counting from
the lowest frequency) in this series are even stronger than the 414~s
peak. In the second power spectrum this signal is completely absent.

The only identical peaks in both power spectra are at the above
mentioned peaks at 0.049 Hz (slightly stronger in the first power
spectrum) and 0.045 Hz (slightly stronger in the second one) and one
futher peak at at $5.26 \times10^{-3}$ Hz (190~s).

As a further method to analyse the significance of the peaks in the
power spectrum we used a bootstrapping technique to estimate the
statistical strength of random peaks. Therefore, we randomized the PN
light curve and calculated the power spectrum. In 1000 runs the maxium
peak has a median strength of 0.08 and never reaches higher than
0.16. This means we have a 0.999 confidence level that the 414~s
peak with a strength of 0.26 is real. For the other peaks, in
particular the 21~s and 22~s peaks -- with strengths of 0.085
and 0.07, respectively -- the significance is questionable and is only
supported by the fact that they appear in two independent data sets
(first half and second half of the PN light curve).

Futhermore, we analysed the MOS1 and MOS2 light curve for any
signals. Since the signal-to-noise level is much lower than in the PN
light curve, only the orbital peak and the peak at the 5f overtone of
the 414~s period are well above the significance level. However, while
the 414~s and 190~s peaks are present (even if with low strengths),
the two 21~s peaks are not.

To examine whether there is a significant orbital modulation in the
data set, we binned the PN light curve to 50 bins and fitted sinusoids
of a number of periods to the data set. Fig.~\ref{Figchi2} shows the
resulting reduced $\chi^2$'s of the best fitting sinusoid vs. the
trial period. The minimum of the reduced $\chi^2$ lies at $P_x =
0.89^{\SSS +0.14}_{\SSS-0.10} \Porb$. While the minimum at $P_x$ seems
to be clearly shifted to 90\% $\Porb$, the orbital period lies within
the 1-$\sigma$ error range of $P_x$. The amplitude of the sinusoid is
$0.41 \pm 0.04$ counts s$^{-1}$
and the zero-line lies at $3.89 \pm 0.02$ counts s$^{-1}$.

   \begin{figure}
   \centering
   \includegraphics[width=8cm]{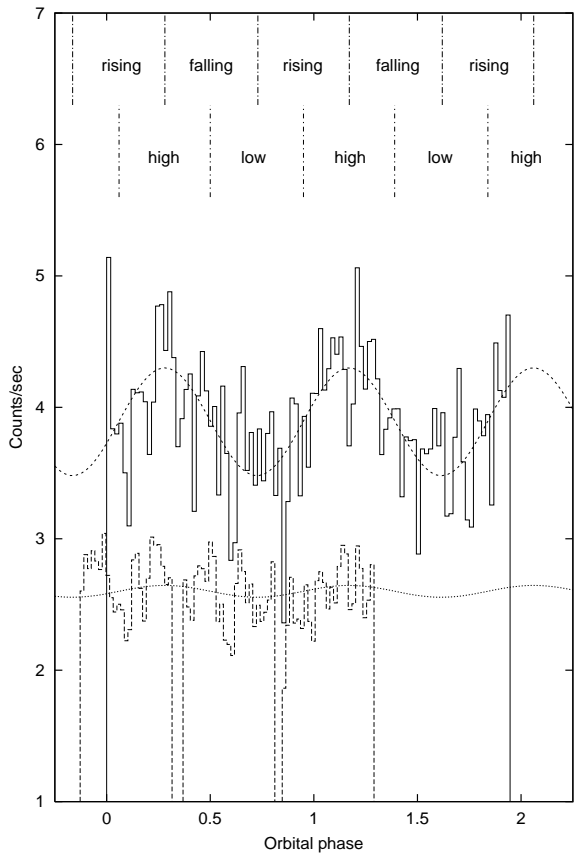}
      \caption{Binned PN and OM (UVW1) light curves (solid and dashed
              histograms, respectively) with sinusoidal
              fit of period 0.89 $\Porb$ to the PN data and
              a scaled sinusoidal function overlayed on the
              OM data. The labels above the light curves are for
	      reference in the text (Section~\ref{explspec} and
	      Fig.~\ref{Figreblhilo}).
              }
         \label{Figpnaom}
   \end{figure}
%

In Fig.~\ref{Figpnaom} we show the binned PN light curve together with
the best sinusoidal fit. The same plot also shows the binned UVW1
light curve. While it is perfectly acceptable to fit the UV data with
a constant count rate ($\chi^2 = 7.20$), an overlayed and scaled
sinusoid (with amplitude 0.045 counts s$^{-1}$
and zero line at 2.6 counts s$^{-1}$)
suggests that the sinusoidal variation seen in
X-rays {\em might} also be present at UV wavelengths ($\chi^2 = 7.06$).

It is noticeable that several of the `down' peaks in the PN and UVW1
light curves seem to coincide, while it is not the case for the `up'
peaks. This {\em might} mean that the emitting flickering sources have
little coherence between its X-ray and UV emission, but are localized
close enough to be occasionally simultaneously eclipsed or absorbed by
non-emitting material.

   \begin{figure*}
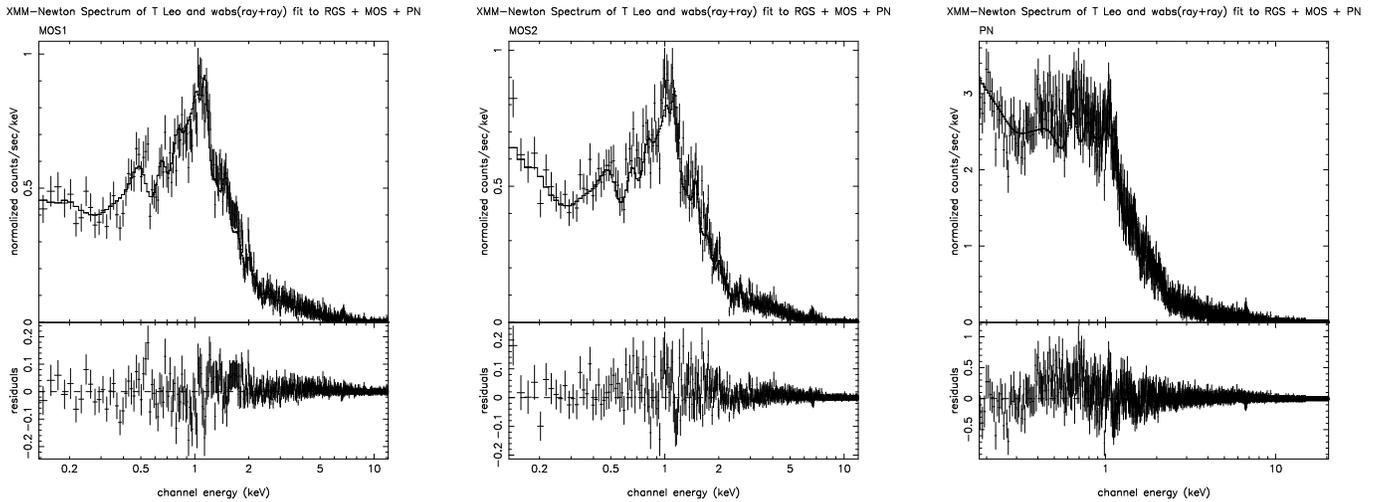


\vspace*{2ex}
   \centering
   \includegraphics[width=5.5cm]{0846f06.ps}
   \hfill
   \includegraphics[width=5.5cm]{0846f07.ps}
   \hfill
   \includegraphics[width=5.5cm]{0846f08.ps}
      \caption{EPIC MOS1, MOS2, and PN spectra with the simultaneous
      fit to the MOS, PN, and RGS data ($\chi^2 = 0.45$). The
      comparison of the individual, observed spectra to the
      model spectra lead to $\chi^2$'s of the MOS1,
      MOS2, and PN of 0.73, 0.66, and 0.44, respectively.
      }
      \label{Figmosfit}
   \end{figure*}
%

\section{Spectroscopy}
\label{spectro}
Using the three different sets of spectra (PN, MOS, and RGS) of
various resolutions and sensitivities, we have a large data set to fit a
spectral model using all data simultaneously. We used {\em xspec} to
fit two $n_H$-absorbed {\em raymond} spectra, following Szkody et al.\
(2001). The fitted parameters are given in Table~\ref{Tabpara}. 
Fig.~\ref{Figmosfit} gives the spectra and fits for the PN and MOS data.

The fitted temperatures are similar to Szkody et al.'s, however, our
column density is much smaller (by a factor of 10) and nearly
vanishing, i.e., at the time of our observations the X-ray emitting
gas was transparent. According to Howell et al.\ (1999) this column
density is identical to that of the interstellar medium in the
direction of T Leo (1.2 to $5 \times 10^{19}$ cm$^{-2}$).

   \begin{table}
      \caption[]{Fitted Parameters for simultaneous fit to MOS, PN, and RGS
      spectra.}
         \label{Tabpara}
     $$ 
         \begin{array}{llll}
            \noalign{\smallskip}
            \hline
            \noalign{\smallskip}
            Parameter{\hspace{4ex}}  &  Value{\hspace{4ex}} & Error{\hspace{4ex}} & Unit{\hspace{4ex}}\\
            \noalign{\smallskip}
            \hline
            \noalign{\smallskip}
            n_H & 0.132  & 0.072 & 10^{20}{\mathrm{cm}}^{-2}     \\
            kT_{{\mathrm{hot}}}  & 4.06  & 0.06 & {\mathrm{keV}}  \\
            Norm_{{\mathrm{hot}}} & 5.396 & 0.010 & 10^{-3}\\
            kT_{{\mathrm{cool}}}  & 0.876  & 0.009 & {\mathrm{keV}}  \\
            Norm_{{\mathrm{cool}}} & 0.240 & 0.041 & 10^{-3}\\
            \noalign{\smallskip}
            \hline
         \end{array}
     $$ Units correspond to those in Szkody et al.\ (2001)
   \end{table}

In the RGS spectra we can identify a few lines as noted in
Table~\ref{Tablines}. However, except for the O~VIII~Ly$\alpha$ line
all lines are too noisy for an analysis of the line profile.  The line
fluxes (counts) were determined with the programme {\it CORA} (Ness \&
Wichmann 2002) using Lorentzian line profiles. The FWHM was in most
cases fixed to 0.04$\AA$.

   \begin{table}
      \caption[]{Lines identified in RGS spectra. The entry ``bad'' means
      that the spectrum has a bad pixel at this wavelength
      ($\lambda$), ``gap'' means the line falls into the gap of the
      spectrum, ``--'' means the line was too faint to be
      measured. Other lines were too faint to determine a total flux.}
         \label{Tablines}
     $$ 
         \begin{array}{lccc}
            \noalign{\smallskip}
            \hline
            \noalign{\smallskip}
            Line{\hspace{4ex}} & {\hspace*{2ex}}\lambda~(\AA){\hspace{2ex}} &
	    {\hspace{2ex}}\mbox{Counts}_{RGS1}{\hspace{2ex}} &
            {\hspace{2ex}}\mbox{Counts}_{RGS2}{\hspace{2ex}} \\
            \noalign{\smallskip}
            \hline
            \noalign{\smallskip}
	    {\mathrm{Mg~XII~Ly}} \alpha & 8.42 & 23 \pm 6 & 19 \pm 6 \\
	    {\mathrm{Fe~XXIII}} & 10.98 & gap & 20 \pm 6 \\
	    {\mathrm{Fe~XVII}} & 11.13 & gap & 23 \pm 6 \\
	    {\mathrm{Ne~X~Ly}} \alpha & 12.13 & gap & 30 \pm 8 \\
	    {\mathrm{Fe~XVII}} & 15.01 & 22 \pm 6 & --\\
            {\mathrm{O~VIII~Ly}} \beta & 16.00 & 23 \pm 6 & 12 \pm 5\\
	    {\mathrm{O~VIII~Ly}} \alpha^1 & 18.97 & 44 \pm 14 & 41 \pm 8 \\
	    {\mathrm{O~VII}} & 21.60 & 8 \pm 4 & gap\\ 
	    {\mathrm{N~VII}} & 24.79 & bad & 14 \pm 5\\
	    {\mathrm{Ar~XIV?}} & 27.55^2 & -- & 10 \pm 4\\
            \noalign{\smallskip}
            \hline
         \end{array}
     $$ 
   $^1$ Counts determined as sum of two Lorentzians fitted to the line
   profile (see Fig.~\ref{Figo8}).\\
   $^2$ blend of Ar XIV 27.47~$\AA$ and Ar XIV 27.63~$\AA$.
   \end{table}

   \begin{figure}
   \centering
   \includegraphics[width=9cm]{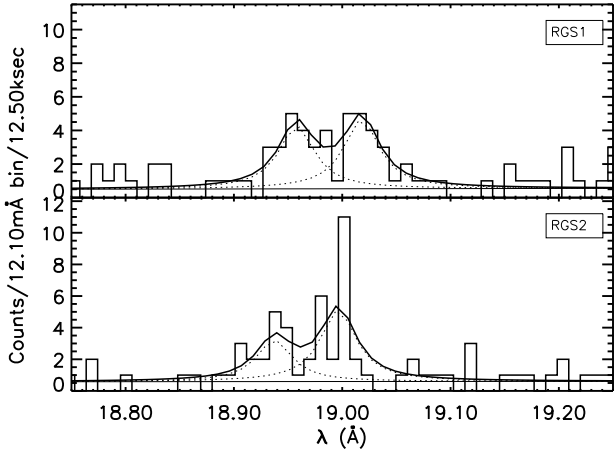}
      \caption{Double peaked O~VIII emission lines in the RGS1 ({\em
   top}) and the RGS2 ({\em bottom}) spectrum (histogram).  The dotted
   lines in each plot are two Lorentzians (FWHM = 0.04\AA), the solid
   line the sum of the two Lorentzians. The $x-$ and $y$-scales in
   both panels are identical.  } \label{Figo8} \end{figure}
%

A more detailed look at the O~VIII line ($\lambda 18.97 \AA$) reveals
a double peaked structure (Fig.~\ref{Figo8}), similar to the emission
lines O~VIII, O~VII, Ne~X, and Ne~IX in the Low Mass X-ray binary
(LMXB) KZ TrA (4U~1626-67) (Schulz et al.~2001). The fitted
Lorentzians have a separation of 950 km s$^{-1}$ in both RGS1 and
RGS2.

As the system ($\gamma$-) velocity of T~Leo was determined to be
$44\pm 20$ km s$^{-1}$ (Shafter \& Szkody 1984) we can ignore it for
our purposes, as the instrumental line broadening is already 0.04\AA
(630 km s$^{-1}$).

In order to determine the significance of the double peaked structure
of the O~VIII line, we fitted the emission line with both single and
double peaks to the line profile in the RGS1 spectrum. Hereby, the
background was set to 25 counts/$\AA$. A fit with instrumental profile
with a width of $0.04\AA$\ leads to a likelihood of ${\cal{L}} =
-80.6$. In comparison, a single Gaussian with $\sigma = 0.042\AA$\
gives an ${\cal{L}} = -89.3$, a single Lorentzian of $FWHM = 0.12\AA$\
leads to ${\cal{L}} = -95.1$. Eventually, a fit of two Lorentzians of
each $FWHM = 0.04\AA$\ gives even a significantly better likelihood of
${\cal{L}} = -96.5$ ($\Delta{\cal{L}} = 1.4$).

Schulz et al.~(2001) explain the double peaked structure with emission
from a narrow, highly ionized layer just above the optically thick
disc surface and below an X-ray heated, fully ionized outer skin of an
accretion disc. Such double peaked emission lines are typical for
accretion discs with high inclination angles. However, the optical
emission lines in T~Leo are not clearly double peaked and the lack of
eclipses points to an inclination angle no larger than 65$^\circ$
(Shafter \& Szkody 1984, Szkody et al.\ 2001).

Inspecting Fig.~\ref{Figo8} it is obvious that the positions of the
O~VIII line in the RGS1 and RGS2 spectra are not identical. While in
the RGS1 the position of the {\it blue} peak is compatible with the
rest wavelength ($\lambda 18.96\AA \pm 0.03\AA$), the {\it red} peak
is at $19.02\AA \pm 0.03\AA$($\sim$ 800 km s$^{-1}$). In the RGS2
spectrum the {\it blue} peak seams clearly shifted shortwards of the
rest wavelength ($\lambda 18.94\AA \pm 0.04\AA$) while the {\it red}
peak is marginally redder than the rest wavelength ($\lambda 19.00\AA
\pm 0.03\AA$). We could exclude an error due to the positioning of the
telescope using the catalogue position for extraction of the
spectra. However, the positions of the red and blue peak in the RGS1
and RGS2 spectra are compatible with each other within the error range
and lead to averages of $\lambda_{\SSS blue} = 18.95\AA\pm 0.03\AA$
and $\lambda_{\SSS red} = 19.01\AA\pm0.02\AA$. In particular, only the
blue peak is compatible with the rest wavelength.  It is however
unclear, how the apparent line shift between the simultaneously taken
RGS1 and RGS2 spectra is caused -- if not by the low photon statistic.

The identification of the O~VII line at 21.6\AA\ in the EPIC MOS1
spectrum allows us to estimate the temperature of the material where
these two oxygen lines are formed from the O~VIII/O~VII line ratio.
This yields a temperature range of $(3$ to $7)\times10^6$~K using the
errors in the total line counts. This temperature is somewhat lower
than the temperature for the cool material found in the fit to all
spectra of $10\times10^6$~K (see Table~\ref{Tabpara}). However,
another cool model component of a temperature as demanded from the
O~VIII/O~VII line ratio would not significantly change the fit to the
spectra (Fig.~\ref{Figmosfit}). Furthermore, we expect to find a
distribution of temperatures in the accretion funnels and the
accretion spots whereas the lower temperatures have much less
significance in the spectral shape.

In the EPIC spectra we can identify a strong Fe XXV line complex. Fitting
a single Lorentzian to this noisy line in the PN spectrum yields a total
flux of $(4.0 \pm 2.1) \times 10^{-5}$ photons cm$^{-2}$~s$^{-1}$
($= (4.2 \pm 2.2) \times 10^{-14}$ erg cm$^{-2}$~s$^{-1}$)
above a rather shallow continuum $(4.3 \pm 3.6) \times
10^{-5}$~photons~keV$^{-1}$~cm$^{-2}$~s$^{-1}$.

\section{Discussion}
\subsection{Origin of the photometric variation}
\label{origin_phot}

   \begin{figure}
   \centering
   \includegraphics[width=8cm]{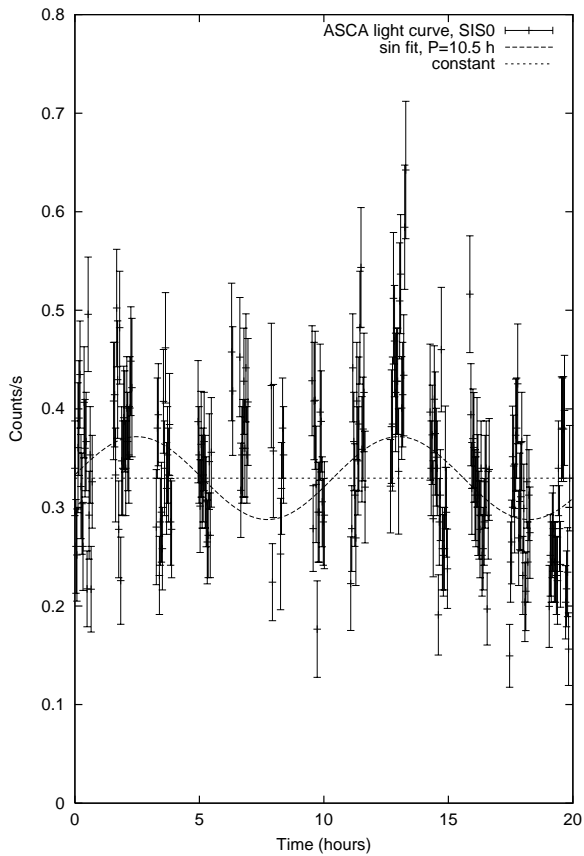}
      \caption{{\it ASCA} SIS0 light curve with overplotted sinusoidal
   fit of period 10.5 h.
              }
         \label{Figasca}
   \end{figure}


We discuss the observed frequencies in the light curve in order
to investigate the nature of the system T~Leo. 

In order to investigate if $P_x$ is caused by a beat period with
$\Porb$, we search for a variation on a timescale of about half a
day. For this purpose we examined {\it ASCA} data from 1998 (Szkody et
al.\ 2001) that were taken over a period of about 20
hours. Fig.~\ref{Figasca} shows that there might indeed be a
sinusoidal variation on a period of $P_y = 10.5$ h. However, the
sinusoid with an amplitude of 0.042 counts s$^{-1}$ has only a
marginally better $\chi^2$ of 2.24 compared to a constant flux
($\chi^2 = 2.62$) and the peak at $\sim 13$~h might also be a flare.

However, apart from the fact that the period $P_x$ has an error
range that is large enough to include the orbital period, it would be
difficult to explain it. Neither a warped disc model (e.g., Patterson
1999) nor a jet model (Livio 1998) can explain the periods $P_x$ or
$P_y$ reasonably well or can be seriously considered. A sudden and
extreme decrease of the orbital period in the last couple of decades
can also be excluded (Kolb 2003, private communication). Therefore, we
consider $P_x$ to be identical to $\Porb$.  For clarity we collected
the definitions of all discussed periods in Table~\ref{Tabperiods}.

In the next two Sections we discuss two models that could explain the
observed period at 414~s: T~Leo is either an intermediate polar or a
dwarf nova showing Quasi-Periodic Oscillations.

   \begin{table*}
      \caption[]{Discussed periods. All but the first period refer to T~Leo.}
         \label{Tabperiods}
         \begin{tabular}{lrl}
            \noalign{\smallskip}
            \hline
            \noalign{\smallskip}
            Symbol{\hspace{1ex}}  & Value & Explanation \\
            \noalign{\smallskip}
            \hline
            \noalign{\smallskip}
	    $P_{\SSS min}$ & $\sim$ 76 min & observed period
            minimum for Cataclysmic variables (e.g.\ Thorstensen et al. 2002)\\
	    $\Porb$ & 84.69936$\pm$0.00068 min & orbital period
            (Shafter \& Szkody 1984) determined 
            from radial velocity measurements of H$\alpha$\\
	    $P_{\SSS sh}$ & 86.7 $\pm$ 0.1 min & super hump period (Lemm
            et al.~1993)\\
	    $P_x$ & $75^{\SSS+12}_{\SSS-8}$ min &
            $0.89^{\SSS+0.14}_{\SSS-0.10} \Porb$ as determined from our
            EPIC PN data, see Fig.~\ref{Figchi2}\\
	    $P_y$ & 10.5 h & determined from {\it ASCA} data, see Fig.~\ref{Figasca}\\
	    $\Pspin$ & 414 s & spin period of white dwarf(?), see Fig.~\ref{Figpower}\\
            \noalign{\smallskip}
            \hline
         \end{tabular}
   \end{table*}



   \begin{figure}
   \centering
   \includegraphics[width=7cm]{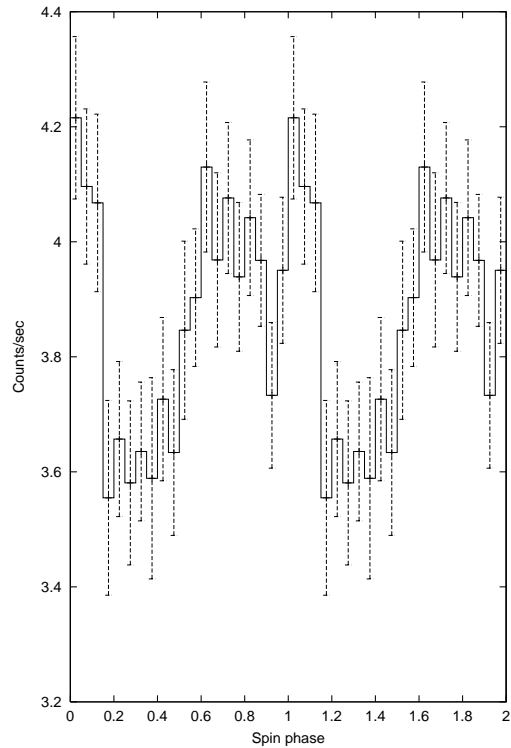}
      \caption{The EPIC PN light curve folded on the period 0.1163~h =
   414.42~s. The error bars are calculated from the scatter in the light
   curve. The light curve is plotted twice for clarity of the plot.
              }
         \label{Figspin}
   \end{figure}
%

\subsubsection{Intermediate Polar model}
\label{spin}
This scenario presumes that the white dwarf is slightly magnetic, just
enough to disrupt the inner part of the disc up
to no more than a few white dwarf radii (however, not to a
complete absence of the disc, as (super) outbursts prove that a disc
must be present). In such an intermediate polar (IP) the accreting
material from the disc is forced to follow the magnetic field lines
onto the two poles on the surface of the white dwarf.  The X-ray
emission then originates from the accretion column close to the white
dwarf (e.g., King 1995).

Such a scenario was already proposed by Shafter \& Skzody~(1984).
While the lack of optical polarisation can be easily explained with a
weak magnetic field, it is more difficult to explain the lack of high
excitation emission lines in the optical range. The lack of obvious
orbital modulation in the optical and UV must then be related to a
relatively low inclination angle of the system whose accretion disc
mainly radiates at these long wavelengths. In contrast, the X-rays
then originate to a large part from the accretion funnels and the
accretion spots on the white dwarf which are much more prone to
viewing angle changes than the disc, even in a low inclination system.
It then seems likely that there is actually a small amount of
variation visible in the {\it UV} as suggested in Fig.~\ref{Figpnaom},
because the accretion funnels will be hot and emit a bluer spectrum,
thus be more pronounced in the UV than in the optical. However, in
both the optical and UV the disc might be considerably brighter than
the accretion funnel.

In IPs the magnetized white dwarf is usually desynchronized with a
spin period of several hundred to thousand seconds. The peak in the power
spectrum of the EPIC PN light curve at 414~s (Fig.~\ref{Figpower})
could thus indeed be interpreted as the spin period of the white
dwarf.
In Fig.~\ref{Figspin} we plot the EPIC PN light curve folded onto the
414~s period. The simultaneous, independent EPIC MOS data are
consistent with the spin profile, however, the shape is less
pronounced due to much higher noise level. In particular, the MOS2
data show a better agreement with the PN data than the MOS1 data.

The spin profile is clearly double peaked with a peak-to-peak
separation of about 0.3 to 0.4 (definitely $< 0.5$) spin phases or
roughly 1/3 of the orbit. Norton et al.\ (1999) observed a similar
situation in the intermediate polar V709 Cas. However, while they see
a peak in the power spectrum at a frequency 3/$\Pspin$, our power
spectrum does not show any obvious peak at this frequency (3/$\Pspin =
7.24\times10^{-3}$~Hz). On the other hand, considering the relative strength
of the 1/$\Pspin$ and 3/$\Pspin$ peaks in Norton et al.'s power
spectrum of about 5:1, a 3/$\Pspin$ signal of an amplitude of the
order 0.025 units might be hidden in the profile of another peak at
$7.17\times10^{-3}$~Hz (see Fig.~\ref{Figpower}).

A more detailed look reveals that in our case the full amplitude of
the variation is only 10\% of the signal (compared to 50\% in V709
Cas). The system is brighter for about 2/3 of the orbit and shows a
sharp drop after the second maximum. This {\em could} indicate a partial
occultation of the X-ray source.

We do not find any significant signal at the beat frequency ($1/\Pspin
- 1/\Porb$), but possibly at ($1/\Pspin - n/\Porb$) with $n = 5,6$,
and at ($1/\Pspin - 1/P_x$) as well as ($1/\Pspin - n/P_x$) with $n =
3,4,6,-1,-2,-3$. As Norton et al.\ (1999) point out, the beat
frequency is usually a signal of stream-fed or disc overflow accretion
and therefore more likely be observed in IPs with stronger magnetic
fields. For T~Leo this means it is one of the few systems that is
disc-fed and seen at a relatively low angle. So far it is unclear what
the observed frequencies represent, if they are significant.

In comparison to other IPs Norton et al.\ (1999) note that such double
peaked profiles appear to arise particularly in white dwarfs with
short spin periods (of less than about 700 s). This appears to be very
well fulfilled. They argue further that the double peaked spin profile
is caused by the small magnetic field (as we propose for T~Leo's white
dwarf). This either leads to large accretion areas and thus broad
accretion curtains with optical depth reversal or (less likely because
of the small inclination angle) accretion columns with conventional
optical depth behaviour but that are tall enough to be partially
visible from behind the white dwarf. In both scenarios the X-ray pulse
maximum is seen whenever one of the poles is closest to the observer
(see also Norton et al.`s Fig.~8).  Since the maxima in the light
curve folded on the spin period are less than 0.5 spin phases apart,
the geometry of the magnetic field must be somewhat asymmetric.

All of this is most striking as the observed variations are only
visible in X-rays, while at optical wavelengths no periodic signal has
been observed (Shafter \& Szkody 1984) and our UV data show only a
very marginal periodicity if at all.

   \begin{figure}
   \centering
   \includegraphics[width=7cm]{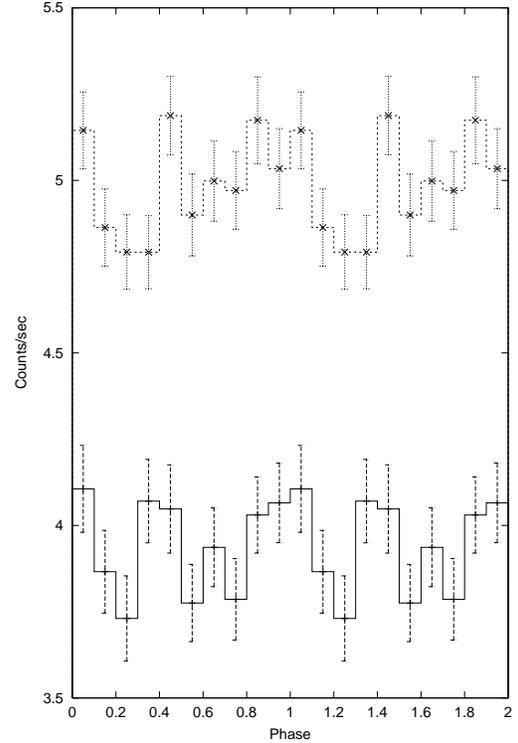}
      \caption{The EPIC PN light curve folded on the period 20.567~s
   (bottom) and 22.112~s (top, shifted upwards by 1 counts s$^{-1}$
   for clarity of the plot).
              }
         \label{Figdno}
   \end{figure}
%

\subsubsection{Dwarf nova model}
\label{QPO}

An alternative model for T~Leo is that it is a dwarf nova showing
Quasi-periodic Oscillations (QPOs) and Dwarf Nova Oscillations (DNOs,
for a review on QPOs and DNOs see Warner 2004). Dwarf novae show often
short-lived QPOs with a period of a few hundred seconds. In
DNO-related QPOs one can typically see DNOs with a period ratio
$P_{\SSS QPO}/P_{\SSS DNO} \sim 15$. The frequencies around 21~s -- if
significant-- could thus actually be the DNOs related to a 414~s QPO.
As the nominal period deduced from the Keplerian velocity at the
radius of the white dwarf is 16~s, it would mean that either the
20.567-s-signal arises from a disc annulus very close to the white
dwarf or that the white dwarf has a high rotational velocity of 800 km
s$^{-1}$.

A test for phase shifts between the first and second half of the data
set does not help in deciding in favour of the QPO model as the
measured phase shift of 0.1 is compatible with zero within the error
range.  Problematic for this model is, however, that the lightcurve
folded on any of the two ``DNO'' periods reveals a double peaked curve
instead of the expected highly sinusoidal signal (Fig.~\ref{Figdno}),
although there might be exceptions (Warner 2004).

However, a number of dwarf novae have been found to show rapid
oscillations in X-rays in quiescence usually of the order a few tens
to a hundred seconds (Table 2 in Warner 2004). In particular, WZ~Sge
is the only one showing persistent DNOs and QPOs in quiescence (Warner
\& Woudt 2002).

The question arises, whether T~Leo is similar to WZ~Sge. T~Leo
certainly shows large amplitude outbursts, but is not generally
considered a candidate for a WZ~Sge type dwarf nova as it also shows
normal outbursts (Kato, Sekine \& Hirata 2001).  WZ~Sge shows DNOs at
27.868~s and 28.952~s (mostly one of the two, but occasionally both
simultaneously) and a period close to the beat period of the two DNOs
at 742~s. These periods are explained as radiation from the central
source with a rotation period of 27.868~s and a thickened region of
the disc with a prograde rotation of period 744~s, the beat period of
the DNOs, that reprocesses and obscures the radiation from the white
dwarf (Warner \& Woudt 2002).

In trying to apply this model to T~Leo, we first notice that close to
the expected beat period of the 20.567~s and 22.112~s periods (294~s)
is only a minor peak with a maximum at 297~s in the power spectrum. We
therefore dismiss the idea that T~Leo displays DNO-related QPO and
consider it likely that the frequencies around 21~s are not significant.

A more likely model, however, is that the QPOs in unrelated to
any DNOs.  One option is that the QPO is either caused by a
modulated mass transfer rate through the inner Lagrangian point due to
oscillations of the secondary similar to the 5-min oscillations on the
sun (Warner 2004).

Another possibility is that the source of the variation is in the
outer disc (Warner 2004). The period of 414~s corresponds to the
Keplerian velocity of the disc material at a distance of $9\Rwd$.
However, as T~Leo is a SU~UMa type dwarf nova and has therefore a
rather large disc, this distance is likely at an intermediate radius,
rather than at the disc edge (the 3:1 resonance radius $R_{3:1}$ is at
$22\Rwd$, see Section~\ref{ring} for system parameters).

We can better imagine a model in which a travelling front
originates close to the inner disc edge of a slightly disrupted
disc. In this scenario, proposed by Warner \& Woudt (2002), the white
dwarf has a low magnetic field, just large enough to disrupt the very
inner part of the accretion disc. The stream from the secondary
overflows the disc and impacts onto the inner disc edge. This can
excite a buldge travelling retrogradely in the rotating disc (in fact
a low frequency prograde $m=1$ g-mode). Using Warner \& Woudt's eq.~19
and 21 and the system parameters as described in Section~\ref{ring},
the impact radius is between 3.1 and 3.8 $\Rwd$ and the Keplerian
period at this radius is between 160~s to 205~s (for mass ratios
$q=1/4$ to $1/6$). Thus, if the buldge is close to the disc edge it
has a speed between a third to half the velocity of the disc
material.

\subsection{Origin of the emission line profile}

\label{explspec}
The broadened, double peaked emission line we observe for O~VIII
could be caused by
a) an accretion disc or ring, similarly to double peaked optical line
   emission in high inclination systems;
b) a spot on the disc edge of an otherwise X-ray dark disc; in the
dwarf nova model or
c) accretion curtains in which the matter from the inner disc radius
   falls down onto a magnetic white dwarf, in the intermediate polar
   model.

In the following section we investigate which of these scenarios apply
to T~Leo.

\subsubsection{Accretion Disc or Ring}
\label{ring}

The explanation that the O~VIII emission arises in an accretion disc
seems unlikely since the optical emission lines show only a very
marginal double peaked structure (Shafter \& Szkody 1984),
particularly not as clear as the separation in the O~VIII line. The
inclination angle is with $i<65^\circ$ (Shafter \& Szkody 1984, Szkody
et al.\ 2001) too small to produce double peaks.  It is, however,
noticeable that the separation of the peaks in the O~VIII line of $\sim$950
km s$^{-1}$, i.e., $\pm475$ km s$^{-1}$, fits to the velocities in the
accretion disc in the Doppler Images of Shafter \& Szkody.

It is, therefore, worth investigating if the emission in O~VIII might
be originating in a ring of narrow radial width in the disc. Simple
model calculations can produce a double peaked U-shaped profile. While
the model profiles have significant flux at the system velocity of the
line, the observed profiles appear more clearly separated. However,
due to the large noise in the data, we cannot exclude that the
emission line profiles are produced in this way.

For a Keplerian ring in the accretion disc we can estimate the radius
from the velocity $v \sin i$ and the white dwarf mass $\Mwd$. The
system parameters of T~Leo are quite uncertain, various values can be
found in the literature. According to Shafter \& Szkody (1984) the
white dwarf has a mass $\Mwd < 0.4 \Msol$ and an inclination angle $i
< 65^\circ$. This leads to an upper value for the ring radius of $0.31
\Rsol \sim 31 \Rwd$ (assuming $\Rwd = 0.01 \Rsol)$. The values
derived by Belle et al. (1998) from model fitting of an IUE spectrum
during an outburst of $\Mwd = 0.6 \Msol$ and $i = 40^\circ$ give a
ring radius of $0.23 \Rsol \sim 20 \Rwd$ using their $\Rwd$ of
$8\times10^8$ cm. 

Using a secondary mass of $M_2 = 0.1 \Msol$ (period-mass relation,
Warner 1995) we obtain a mass ratio of $q = \frac{1}{4}$ or
$\frac{1}{6}$, respectively. This leads to a distance from the white
dwarf to the inner Lagrangian point $L_1$ in relation to the binary
separation $a$ of $\Rl/a = 0.65 \pm 0.02$ (determined as the average of
the $\Rl/a$'s derived for the two values of $q$; Silber 1992, after
Warner 1995). Using Warner's (1995) mass-radius relationship for the
secondary gives a secondary radius of $R_2 \sim 0.136 \Rsol$. With
this we can derive the absolute distance from the white dwarf to the
inner Lagrangian point $\Rl$ to $0.25 \Rsol$ or 22$\Rwd$, i.e., for
the following 
discussion we use Belle et al.'s (1998) system parameters as their
parameters allows the proposed ring to be inside the Roche-lobe of the
white dwarf. However, the true value must be somewhat larger, as the
3:1 resonance radius $R_{3:1}$ is at also $22\Rwd$, and during super
outbursts the disc becomes larger than this. The true system parameters
must therefore be slightly different from the quoted values.

   \begin{figure} \centering
   \includegraphics[width=8cm]{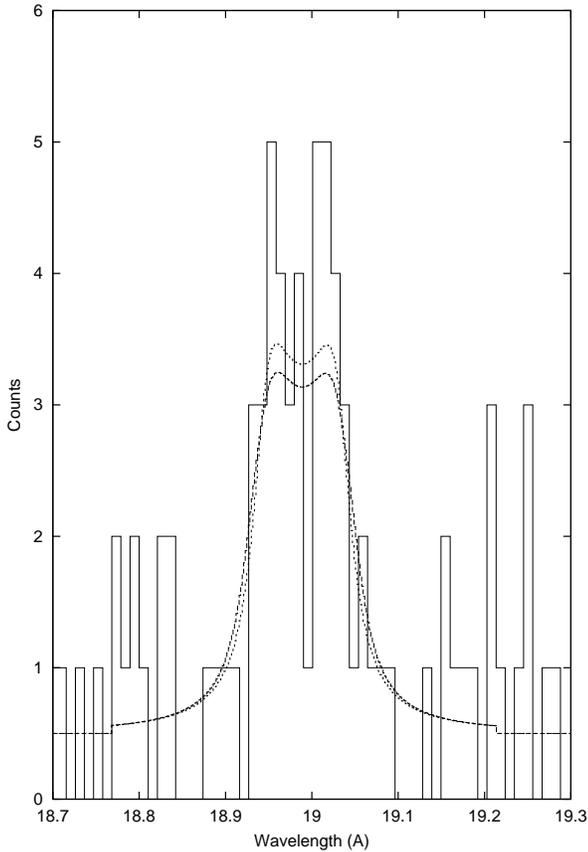} \caption{O~VIII
   emission line with overplotted models for an emitting ring in the
   accretion disc. The model profiles were convolved with a Lorentzian
   of $0.04\AA$ width. The ring is either narrow at a radius of
   $8\Rwd$ ($0.37\Rl$, dashed line) or has a radius range between
   maximally $4\Rwd$ to $14\Rwd$ ($0.18$ to $0.65\Rl$, dotted
   line). The normalization of the model profiles are chosen simply
   for illustrative purposes.} \label{Figprof} \end{figure}


As an example we constructed a simple model of an emitting ring and
calculated the emission line profile folded with a Lorentzian with the
observationally given resolution of $0.04\AA = 630$ km s$^{-1}$. We
could only achieve a double peaked profile with the peaks at the
observed velocities ($\pm 475$ km s$^{-1}$) and the observed line
width for either a very narrow ring ($< 1$\%) at about $8\Rwd$
($0.37\Rl$) or a broader ring between maximal $4\Rwd$ and $14\Rwd$ ($0.18$ to
$0.65\Rl$) (Fig.~\ref{Figprof}). However, it is impossible to
separate the two peaks clearly by assuming a ring structure.

This calculation shows that the ring model does not convincingly
reproduce the observed results. Furthermore, the ring would have to be
quite far from the hot inner disc ($>4\Rwd$) and it is unclear why
only there the O~VIII emission would be produced, as O~VIII has a
maximum at a temperature of $3\times10^6$~K. Such temperatures are
only expected for the inner part of the disc (i.e., very close to the white
dwarf).


In principle we can save this idea, if we assume that the disc is {\em
warped} in the inner part of the disc. This would mean that the hot,
inner disc has a different and variable inclination angle than the
rest of the disc. This could explain the lack of correlation between
the gradients in the light curve and the presence of double peaks as
well as the single peaked optical emission lines. When we see it edge
on, it becomes double peaked, however, at times when we face the inner
part of the disc, the emission line becomes {\it narrow} not broad.
However,
this scenario is also problematic.

   \begin{figure*} 
   \centering
   \includegraphics[width=8cm]{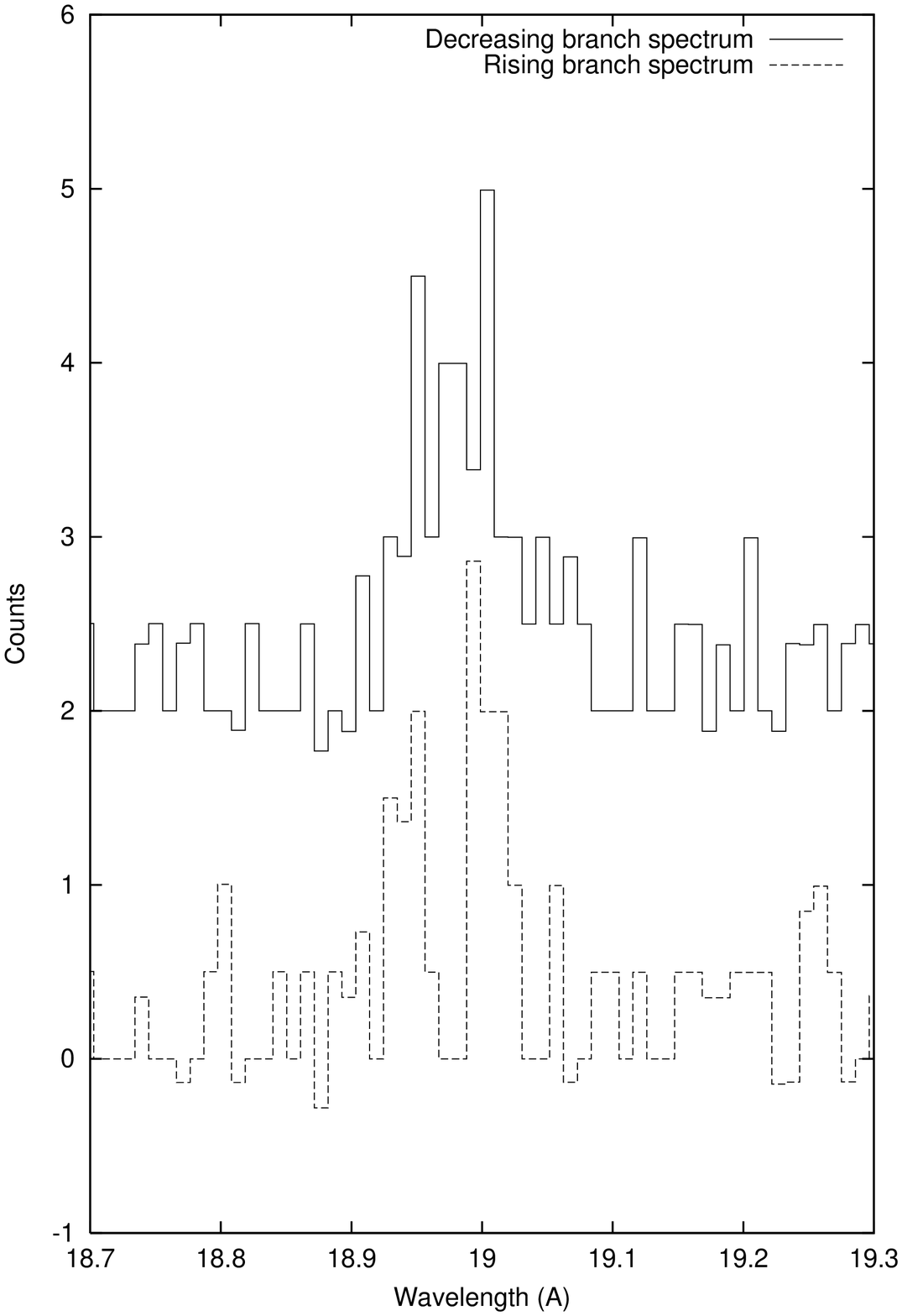}
   \includegraphics[width=8cm]{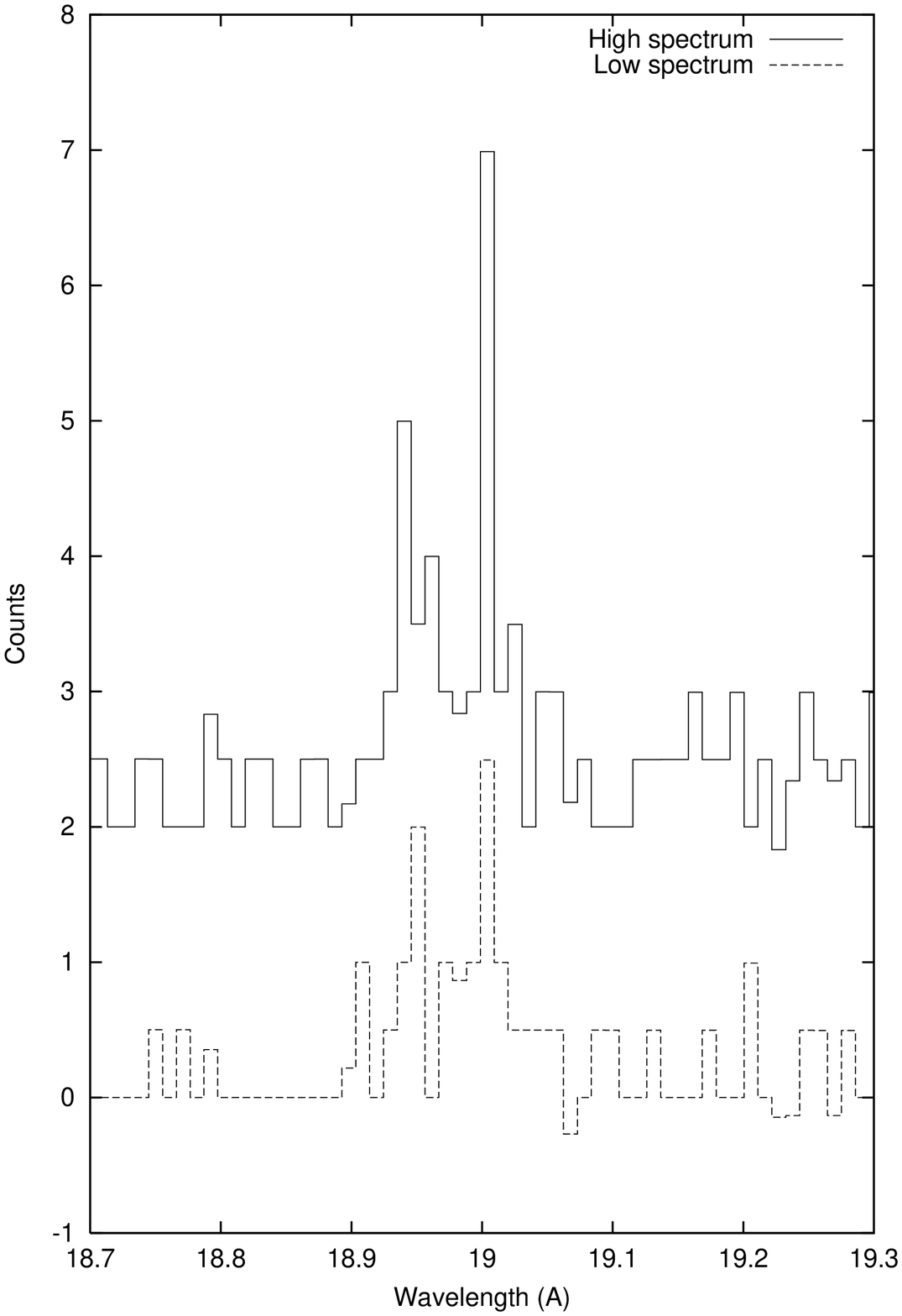}

   \caption{Spectra during rising and falling (decreasing) phases
   ({\em left}) and spectra during high and low phases ({\em
   right}). The spectra for the decreasing branch phases and the
   high phases are shifted upwards by 2 counts for clarity of the
   plots. See Fig.\ref{Figpnaom} for identification of phase ranges.}

   \label{Figreblhilo}
   \end{figure*}


\subsubsection{X-ray bright hot spot}

In order to investigate option b), we compared the timings of the
spectra with those of the X-ray light curve. Hereby, we assume that
the sinusoidal variation in the PN light curve is caused by the bright spot
moving in and out of view. Then we can separate the light curve in a {\em
falling} portion, when the X-rays fade, and a {\em rising} portion,
when the X-ray brighten according to Fig.~\ref{Figpnaom}. If the
beaming hot spot theory is correct, we would expect that the spectra
produced from timings of the {\it falling} portion of the light curve
only show a red peak in the O VIII emission line, while the other
spectra created from timings of the {\it rising} portions of the light
curve should show only a blue peak. However, this is not the case
(Fig.~\ref{Figreblhilo}, left panel). While the ``{\it blue}''
spectrum clearly shows a double peaked structure for the O VIII line,
the ``{\it red}'' spectrum simply shows a broadened emission line.
Therefore, we have to dismiss this scenario.

\subsubsection{Accretion curtains in an Intermediate Polar}

Eventually, we investigate, if (and how) the double peaked
emission line can be produced within an IP model. This may
include a scenario in which the white dwarf has a very low field as
proposed for the QPO model (end of Section~\ref{QPO}).
This goes hand in hand with the question, if thus we
should expect such {\em emission line profiles} for all IPs with small
magnetic fields (and double peaked {\em spin profiles}).

In an IP the matter from the secondary is first fed into an accretion
disc. This disc is disrupted at a certain inner radius from where the 
matter couples to the magnetic field lines of the white dwarf. This
leads to accretion curtains through which the matter from the
secondary is dumped onto the surface of the white dwarf.

Whenever the
magnetic north pole (by definition the pole that is on the hemisphere
closer to the observer) is facing the observer, the material in the
accretion curtain is falling down onto the white dwarf, causing red
shifted emission. However, at the timing of the second maximum the
white dwarf has turned such that the material in the curtains moves
perpendicular to the line of sight.  Even if the accretion columns are
tall enough so that the southern one is visible at the time of the
first maximum from behind the white dwarf we would not expect
blue-shifted emission, as the X-rays are produced too close to the
white dwarf to be visible and the curtain bends too strongly to see any
emitting material falling in the direction towards the observer.

We can expect both blue and redshifted emission, if the poles are
moving in and out of view due to the rotation of the white
dwarf. However, the rotational velocity of the white dwarf (using
Belle et al.'s (1998) radius of $8\times10^8$~cm) is $2\pi \Rwd /
\Pspin = 121$~km s$^{-1}$, much less than the observed $\pm 475$
~kms$^{-1}$ in the emission line profile.

As mentioned in Section~\ref{spectro} emission line peaks are not
shifted symmetrically to the rest wavelength, where the blue peak is
compatible with the rest wavelength and the red one shifted to
a positive velocity
of about $v = +630\pm340$ km s$^{-1}$. This means, we see the red
peak originating in the accretion funnel of the northern hemisphere in
which the material is moving away from the observer onto the white
dwarf (involving an optical depth reversal as proposed by Norton et
al.~1999). The blue peak is not seen, as the infalling material with
velocity towards the observer is hidden from the view behind the
white dwarf. The peak at the rest wavelength comes from material
moving perpendicular to the line of sight, e.g., at the phase when the
southern pole is closest to the observer.

An analysis of spectra extracted during the phases of the two spin
profile maxima and the single minimum ($0.5-0.925, 0.925-1.15,
0.15-0.5$, respectively, see Fig.~\ref{Figspin})
seems to support this idea (e.g.\ the emission during the broad
maximum between spin phases 0.5 to 0.925 appears to show only the red
peak), however, the count rates are too low to make any firm
statements. Further observations with a more secure wavelength
calibration and better S/N ratio are necessary to confirm this model.

The IP model for T~Leo was first suggested by Shafter \& Szkody
(1984). In order to explain their measured radial velocity curve shift
Warner (1995) proposes that the accretion stream from the secondary
overflows the accretion disc and hits the inner disc edge similarly as
in EX~Hya. Shafter \& Szkody's observed line width of H$\alpha$ in
fact is compatible with an inner disc radius of 6-7 $\Rwd$. However,
as the disc is much cooler than the acretion funnels we do not expect
any implication of the stream overflow model on our X-ray data.

T~Leo is not the only suspect for an SU~UMa type dwarf nova showing
characteristics of an IP, however, it is the strongest candidate.
SW UMa also shows superoutbursts with the typical features like superhumps
(Robinson et al. 1987) and Shafter, Szkody \& Thorstensen (1986)
discovered a period of 15.9 min that could be the rotation period of
the magnetic white dwarf. However, Rosen et al.\ (1994) could not
confirm the 15.9~min period with their data.

While the superoutbursting (Kato \& Nogami 1997) VZ Pyx has been
suspected to be an IP (de Martino et al.\ 1992, Remillard et
al. 1994), Szkody \& Silber (1996) point out, this system might not be
an IP, as it shows coherent pulses only during decline from
outburst. In contrast, our spin detection was made outside of any
outburst. Furthermore, Warner, Woudt \& Pretorius (2003) find QPO and
DNO signals in their photometry of VZ~Pyx, excluding the possibility
that it is an IP. Another candidate is HT Cam (RX J0757.0+6306) which shows
oscillations (Kemp et al.\ 2002), but whose SU~UMa nature is not
confirmed, yet (Tovmassian et al.~1998).

Further objects listed in Warner's (1995) Table 7.2 being SU~UMa type
dwarf nova and simultaneously IP candidates are AL~Com and
HT~Cas. However, while both are clearly SU~UMa stars showing super
outbursts, there is no clear evidence that HT~Cas is an IP
(but possibly a VY Scl type object, Robertson \& Honeycutt 1996), and
AL~Com shows only some properties similar to EX~Hya (Abbott et
al. 1992). All in all both are rather unusual objects, possibly linked
to the WZ~Sge-type dwarf novae.

\section{Summary}

The X-ray light curve reveals a period at 414~sec that could be the
spin period of the white dwarf $\Pspin$.  Proven correct, this would
make T~Leo the first confirmed intermediate polar that displays
superoutbursts. In this model it is relatively easy to explain the
double peaked X-ray emission line O~VIII (Ly $\alpha$) originating in
the accretion curtains. Furthermore, it would make T~Leo extremely
interesting, as it provides a unique opportunity to investigate the
disc instability in a magnetic cataclysmic variable.

An alternative model is that the 414~sec signal is a Quasi-periodic
Oscillation, leaving T~Leo to be a dwarf nova as usually assumed. 
These QPOs arise most likely in a mass transfer variation from the
secondary or a distortion in the intermediate regions in the disc.
However, it is very difficult -- if not impossible -- to explain the
double peaked emission line O~VIII (Ly $\alpha$) in this model as the
optical lines are single peaked.

It is, therefore, highly desirable to obtain further X-ray high
resolution photometry and spectroscopy to investigate the stability of
the 414~sec signal and the shape of the X-ray emission line
profiles. Only this will give us unambiguous answers concerning the nature
of T~Leonis.

\begin{acknowledgements}
This work is based on observations obtained with {\it XMM-Newton}, an
ESA science mission with instruments and contributions directly funded
by ESA Member States and the USA (NASA).  S.V. and J.-U.N. acknowledge
support from DLR under 50OR0105. Furthermore, we thank the referee for
valuable comments leading to an improvement of the paper.
\end{acknowledgements}

\end{document}